\documentclass[namedreferences]{solarphysics}

\usepackage[hyperref,optionalrh]{spr-sola-addons} 
\usepackage{graphicx}        
\usepackage{color}           
\usepackage{breakurl}        

\chardef\us=`\_
\begin{document}

\begin{article}
\begin{opening}

\title{Development of Fast and Precise Scan Mirror Mechanism for an Airborne Solar Telescope}

\author[addressref={aff1},corref,email={takayoshi.oba@nao.ac.jp}]{\inits{T.}\fnm{Takayoshi}~\lnm{Oba}\orcid{0000-0002-7044-6281}}
\author[addressref={aff2}]{\inits{T.}\fnm{Toshifumi}~\lnm{Shimizu}\orcid{0000-0003-4764-6856}}
\author[addressref={aff1}]{\inits{F.}\fnm{Yukio}~\lnm{Katsukawa}\orcid{0000-0002-5054-8782}}
\author[addressref={aff1}]{\inits{F.N.}\fnm{Masahito}~\lnm{Kubo}\orcid{0000-0001-5616-2808}}
\author[addressref={aff1}]{\inits{F.N.}\fnm{Yusuke}~\lnm{Kawabata}\orcid{0000-0001-7452-0656}}
\author[addressref={aff1}]{\inits{F.N.}\fnm{Hirohisa}~\lnm{Hara}\orcid{0000-0001-5686-3081}}
\author[addressref={aff1}]{\inits{F.N.}\fnm{Fumihiro}~\lnm{Uraguchi}}
\author[addressref={aff1}]{\inits{F.N.}\fnm{Toshihiro}~\lnm{Tsuzuki}\orcid{0000-0002-8342-8314}}
\author[addressref={aff1}]{\inits{F.N.}\fnm{Tomonori}~\lnm{Tamura}}
\author[addressref={aff1}]{\inits{F.N.}\fnm{Kazuya}~\lnm{Shinoda}}
\author[addressref={aff3}]{\inits{F.N.}\fnm{Kazuhide}~\lnm{Kodeki}}
\author[addressref={aff3}]{\inits{F.N.}\fnm{Kazuhiko}~\lnm{Fukushima}}
\author[addressref={aff4}]{\inits{F.N.}\fnm{Jos\'{e} Miguel}~\lnm{Morales Fern\'{a}ndez}\orcid{0000-0002-5773-0368}}
\author[addressref={aff4}]{\inits{F.N.}\fnm{Antonio}~\lnm{S\'{a}nchez G\'{o}mez}}
\author[addressref={aff4}]{\inits{F.N.}\fnm{Mar\'{i}a}~\lnm{Balaguer Jimen\'{e}z}}
\author[addressref={aff5}]{\inits{F.N.}\fnm{David}~\lnm{Hern\'{a}ndez Exp\'{o}sito}}
\author[addressref={aff6}]{\inits{F.N.}\fnm{Achim}~\lnm{Gandorfer}}

\address[id=aff1]{National Astronomical Observatory of Japan, 2-21-1 Osawa, Mitaka, Tokyo 181-8588, Japan}
\address[id=aff2]{Institute of Space and Astronautical Science, Japan Aerospace Exploration Agency, 3-1-1 Yoshinodai, Chuo, Sagamihara, Kanagawa 252-5210, Japan}
\address[id=aff3]{Mitsubishi Electric Corporation, 8-1-1 Tsukaguchihonmachi, Amagasaki, Hyogo 661-8661, Japan}
\address[id=aff4]{Instituto de Astrof\'{i}sica de Andaluc\'{i}a, Glorieta de la Astronom\'{i}a s/n, 18008 Granada, Spain}
\address[id=aff5]{Instituto de Astrof\'{i}sica de Canarias, C. V\'{i}a L\'{a}ctea, s/n, 38205 San Crist\'{o}bal de La Laguna, Santa Cruz de Tenerife, Spain}
\address[id=aff6]{Max-Planck-Institut f\"{u}r Sonnensystemforschung, Justus-von-Liebig-Weg 3, 37077 G\"{o}ttingen, Germany}

\runningauthor{T. Oba et al.}
\runningtitle{\textit{Solar Physics} Development of Scan Mirror Mechanism }

\begin{abstract}
We developed a scan mirror mechanism (SMM) that enable a slit-based spectrometer or spectropolarimeter to precisely and quickly map an astronomical object. 
The SMM, designed to be installed in the optical path preceding the entrance slit, tilts a folding mirror and then moves the reflected image laterally on the slit plane, thereby feeding a different one-dimensional image to be dispersed by the spectroscopic equipment. 
In general, the SMM is required to scan quickly and broadly while precisely placing the slit position across the field-of-view (FOV). 
These performances are highly in demand for near-future observations, such as studies on the magnetohydrodynamics of the photosphere and the chromosphere. 
Our SMM implements a closed-loop control system by installing electromagnetic actuators and gap-based capacitance sensors. 
Our optical test measurements confirmed that the SMM fulfils the following performance criteria: i) supreme scan-step uniformity (linearity of 0.08$\%$) across the wide scan range ($\pm 1005^{\prime \prime}$), ii) high stability ($3 \sigma=0.1^{\prime \prime}$), where the angles are expressed in mechanical angle, and iii) fast stepping speed (26 ms). 
The excellent capability of the SMM will be demonstrated soon in actual use by installing the mechanism for a near-infrared spectropolarimeter onboard the balloon-borne solar observatory for the third launch, \textsc{Sunrise III}.
\end{abstract}
\keywords{solar physics, mirror, tip-tilt, chromosphere, photosphere}
\end{opening}

\section{Introduction}
     \label{S-Introduction} 
Spectroscopic and spectropolarimetric observations provide measurements for physical quantities such as temperature, velocity, density, and vector magnetic field in the solar atmosphere. 
One of the most commonly used observation system in a spectrograph is a slit-based instrument. 
The slit slices an incoming light that passes through it into a one dimensional image. 
Then a spectroscopic instrument, such as a grating, disperses this one-dimensional image in a direction perpendicular to the slit, so that a camera captures a two-dimensional image that consists of spatial and wavelength directions. 
While this optical system suffers from the limitation of recording only spatially one-dimensional data at a time, a scan mechanism preceding the slit displaces an image laterally on the slit plane, thereby dispersing another one-dimensional image that can be captured by the camera. 
A repetition of this tip-tilt process produces many one-dimensional slices at different times, creating a spatially two-dimensional spectral map. 
In such a slit-based observation instrument, a reliable  mechanism (hereinafter called the scan mirror mechanism or SMM in an abbreviated form) is necessary to precisely map the two dimensional distribution of the physical quantities inferred from the spectroscopic or spectropolarimetric data. \\
The SMM is generally required to deliver the following fundamental functionalities: wide coverage of the scan range, fast scan stepping, and high stability of angle positioning. 
These functionalities are increasingly in demand in recent years, especially for studies on magnetohydrodynamics of the photosphere and chromosphere. 
In these atmospheres, each magnetic element is distributed on a fine scale (a few hundred km or smaller), but many of them are intermingled with each other and form network structure on a large scale (a few tens of megameters). 
Furthermore, they are highly dynamic, since a number of magnetohydrodynamic waves continuously propagate across the network and sporadically induced magnetic reconnection may change the topology of the network structure itself. \\
In recent decades, several types of the slit-scan systems have been developed and adopted in space-borne solar telescopes. 
One of the most commonly adopted systems is a stepper motor that drives a ball screw. 
This type of system is implemented in the Coronal Diagnostic Spectrometer (CDS: \citealt{Harrison1995}) onboard the Solar and Heliospheric Observatory (SOHO), and the visible spectropolarimeter \citep{Lites2013b, Shimizu2008} onboard the Hinode satellite \citep{Kosugi2007}. 
While offering high-resolution positioning, the stepping motor system takes a long time to complete the flyback motion, which is a necessary process for repeated scanning, as it moves back from the end position of the scan to the initial position. 
In the case of the Hinode/Spectropolarimeter, it takes longer than 10 minutes to flyback over the full scan range. 
Another commonly used system is a piezoelectric transducer. 
This type of system is adopted in the slit-scan system onboard the IRIS spacecraft \citep{DePontieu2014}, the EUV imaging spectrometer (EIS) onboard the Hinode satellite \citep{Culhane2007}, and the Spectral Imaging of the Coronal Environment (SPICE) instrument onboard the Solar Orbiter satellite \citep{SpiceConsortium2020, Muller2020}. 
While benefitting from inducing a quick motion, the piezoelectric actuator encounters a drawback of hysteresis.
The property of hysterisis is that the mirror angle position differs depending on its movement in the past, and thus the absolute positioning is not assured. 
The most effective approach to eliminate hysteresis is to use a closed-loop control. 
In this control, the actual tilt angle is measured in real time by equipped sensors, e.g., strain gauges in the case of the Hinode/EIS, and fed back to ensure tilting toward the correct position according to the outputs of the sensors. 
In the past, space-borne solar spectrometers had a scan ranges of a few hundred arcsec in mechanical angle, and step speeds of a few hundreds milliseconds. 
We designed an SMM that delivers superior performance, achieving a scan range wider than a thousand arcsec in mechanical angle, fast scan stepping from one position to another within a few tens of ms, and a quick flyback speed of a few tens of ms as well. 
The SMM uses an electromagnetic actuator in order to accomplish a broad stroke with a fast tilt. 
Furthermore, we implemented a closed-loop control into the SMM in order to provide accurate tilt-positioning. \\
The developed SMM is to be installed for a near-infrared spectropolarimeter onboard the balloon-borne solar observatory, \textsc{Sunrise III}. 
This balloon project is being carried out with an international cooperation among Germany, Spain, the United States, and Japan. 
The observatory employs a 1-meter prime mirror, the largest in the airborne solar observatories. 
The flying altitude is 35-37 km, where there is no seeing-induced image degradation. 
This balloon project has experienced two launches in 2009 and 2013 (\citealt{Barthol2011, Solanki2017}). 
In the upcoming third launch scheduled for 2022, in addition to the upgrade of the observation instruments in the previous flights, we have a newly developed \textit{the SUNRISE Chromospheric Infrared spectro-polarimeter}, abbreviated to SCIP (\citealt{Katsukawa2020}). 
This spectropolarimeter covers two bands simultaneously, 770nm and 850nm. 
The SCIP adopts a slit observation system and spans a broad wavelength range. 
Simultaneous observation of multiple lines will measure the physical quantities seamlessly from the photosphere to the chromosphere \citep{Quintero_Noda2017, Quintero_Noda2017b, Quintero_Noda2018}. 
The SCIP is designed to achieve a diffraction-limited spatial resolution of 0.2$^{\prime \prime}$ even in harsh flight conditions, which is ensured by the elaborately designed optical bench \citep{Uraguchi2020}. \\
The first requirement imposed on the SMM is good scan-step uniformity (linearity) across a broad field-of-view (FOV). 
Good linearity ensures a constant spacing in the scan direction of a spectral map. 
In the slit-scan system of the Hinode/Spectropolarimeter, non-equidistant step performance due to a ball-screw mechanism has been reported in the optical measurements carried out prior to the launch \citep{Lites2013b}, and also investigated in onboard data by comparing the scanned-map with the filtergraph images \citep{Centeno2009}. 
The second one is the high-stability of the angle positioning, as it can retain high imaging performance. 
The third one is fast stepping speed, as it can bring the scanned-map closer to an image as if it were taken instantaneously and improve the time cadence of the scan repetition. 
Furthermore, the SMM should be operated nominally and retain its scan function under the severe temperature and pressure environment during the balloon flight in the stratosphere. \\
This paper introduces the mechanical design of the SMM and the overall concept of its scan function, executed together with the related electric components. 
The latter sections describe a series of optical measurement tests and environmental tests that reasonably verify the above-mentioned performances.

\section{SMM (Scan Mirror Mechanism)} 
      \label{S-aug}
\subsection{General description} 
\begin{figure}    
\centerline{\includegraphics[width=0.8\linewidth,clip=]{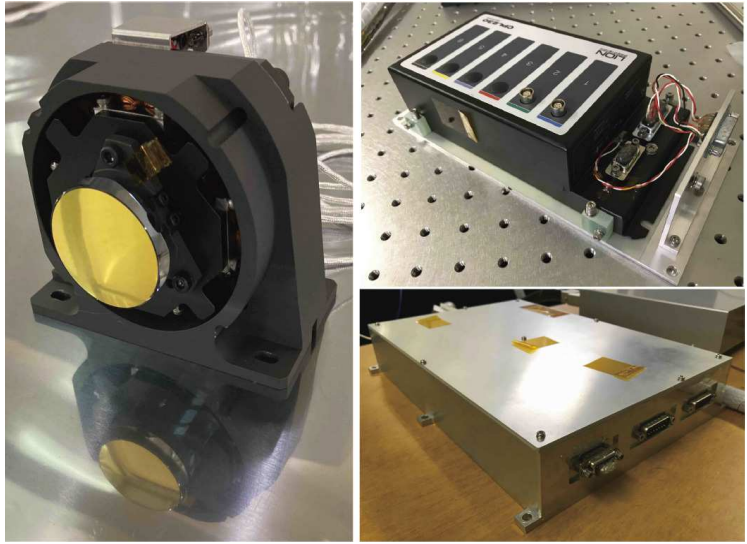}}
\caption{Photographs of the SMM components. SMM-TM attached with the flight mirror with gold coating (left). SMM-IF (upper right). SMM-DRV (lower right).}
\label{fig:photo_smm}
\end{figure}

The SMM consists of three components: a tip-tilt mechanism (SMM-TM), electronics that drive a mirror tilt (SMM-DRV: Driver), and an electrical circuit to precisely read the output signal from gap-based capacitance sensors installed on the SMM-TM (SMM-IF: Interface), as shown in Fig.\ref{fig:photo_smm}. 

\begin{figure}    
\centerline{\includegraphics[width=0.8\linewidth,clip=]{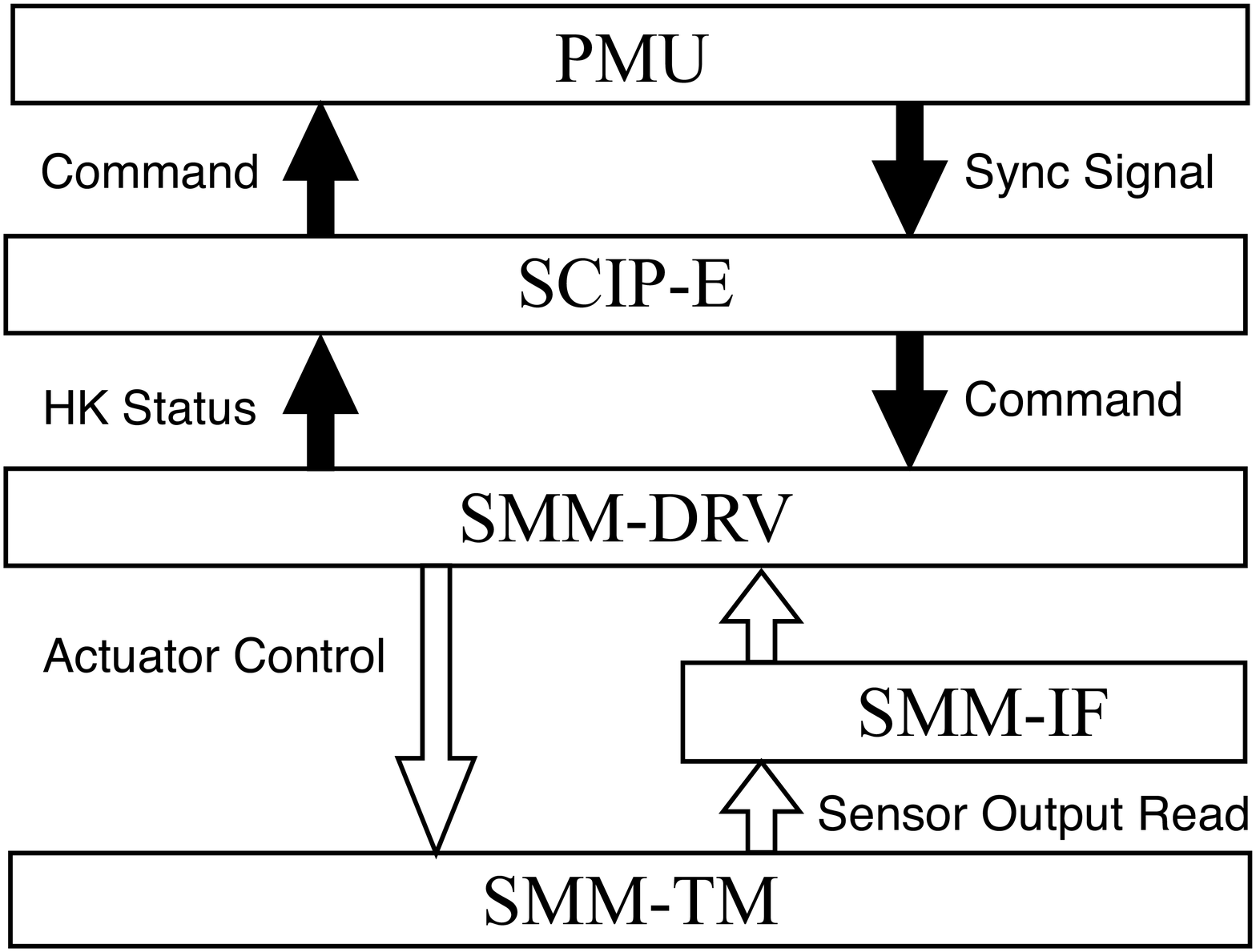}}
\caption{Communication diagram between the SMM-TM, -IF, and -DRV, together with the related components, SCIP-E, and PMU. 
The arrows represent the direction of the data flow. Black filled arrows indicate communication on digital line, whereas the outline arrows indicate communication on analog line. }
\label{fig:comuline}
\end{figure}

Figure \ref{fig:comuline} illustrates the overall configuration of the SMM, showing the electric connections between the three SMM components along with the other electronic instruments. 
All the components are controlled synchronously in the following way. 
The electronics of the SCIP (SCIP-E) receive a synchronization (pulse) signal every 32 ms, generated and transmitted by the PMU (Polarization Modulator Unit: \citealt{Kubo2020}), which rotates a waveplate and modulates the polarized part of the incoming light.
In response to this synchronization signal, SCIP-E transmits a tilt command to SMM-DRV that drives a tip-tilt motion of SMM-TM. 
SCIP-E also transmits a command to activate the camera sensors. 
The camera exposure takes place 32 ms later than the timing of the tip-tilt command, thereby avoiding the activation of the exposure at the mirror movement. 

\begin{figure}    
\centerline{\includegraphics[width=1.0\linewidth,clip=]{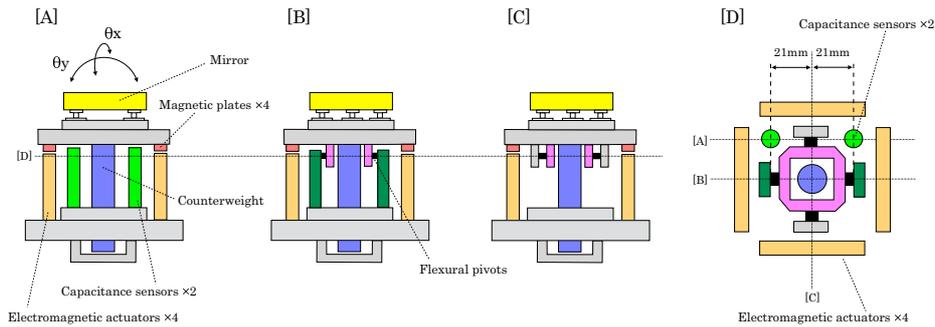}}
\caption{Schematic of the SMM-TM from different view angles. 
The cross section is marked with the dotted line. }
\label{fig:smm_tm}
\end{figure}

Figure \ref{fig:smm_tm} illustrates a schematic of the SMM-TM. 
The electromagnetic attraction system is used for actuating the tilt of the mirror attached to the SMM-TM. 
Four actuators are arranged behind the mirror and placed at 90 degrees from each other. 
Electric current flows through the actuators (coil), thereby inducing a magnetic pulling force that acts on the metal plane mounted behind the mirror. 
One pair of actuators pulls the mirror in one direction, whereas the other pair pulls the mirror in the perpendicular direction. 
$\theta_{y}$-direction is parallel to the scan direction. 
$\theta_{x}$-direction is perpendicular to the scan direction but is used for a coalignment purposes in coordination with other observational instruments. 
A mirror is mounted on a stage supported by two sets of two flexural pivots (5008-400, Riverhawk Corp.) on the orthogonal axes. 
One set is mounted to the bottom structure as depicted in panel [B] of Fig.\ref{fig:smm_tm} and the other set is mounted to the mirror stage as depicted in panel [C], allowing two-dimensional tilt. 
The flexural pivot does not have a sliding portion which makes friction negligible. 
This eliminates the limitation on the practical resolution of mechanical components. \\
The flight mirror is a convex spherical mirror with an effective diameter of 25 mm, mounted on the SMM-TM. 
Since a convex mirror makes tilt measurement difficult with a simple configuration, we accordingly fabricated a flat mirror with a size and weight comparable to the flight mirror. 
The flat mirror can be used for testing since our aim is to inspect the mechanism itself, rather than the optical performance. \\
The SMM-TM is mounted on the relay-optical system of the Image Stabilization and Light Distribution system (ISLiD). 
Its details for the first flight of \textsc{Sunrise}, are described in \citealt{Gandorfer2011}. 
The ISLiD guides the solar image onto the entrance slit of the SCIP. \\

To precisely control a tilt, a closed-loop system is implemented, which requires a feedback architecture to measure the tip-tilt angle in real time. 
Gap-based capacitance sensors (CPL230, Lion Precision) are installed behind the mirror of the SMM-TM. 
The output of the sensors is processed and immediately fed back to the appropriate tilt-position. 
The output of the gap distance is converted to a mirror angle through the tangential relation, $\theta = d/21$, where $\theta$ is a tilt angle, the distance between the center of the mirror and the sensor is 21 mm (see panel \textbf{[D]} in Fig. \ref{fig:smm_tm}), and $d$ is the gap in millimeters with respect to the nominal position. 
$d$ varies over the sensor range of $\pm$0.125 mm in accordance with the tilt angle, which is measured as the output of the gap sensor. 
Two of the gap sensors are arranged in the $\theta_{x}$-direction. 
Thus, to determine $\theta_{x}$, $d$ is taken to be the average of the addition of the outputs of the two sensors, whereas to determine $\theta_{y}$, $d$ is taken to be the average of the corresponding subtraction. \\

\begin{figure}    
\centerline{\includegraphics[width=1.0\linewidth,clip=]{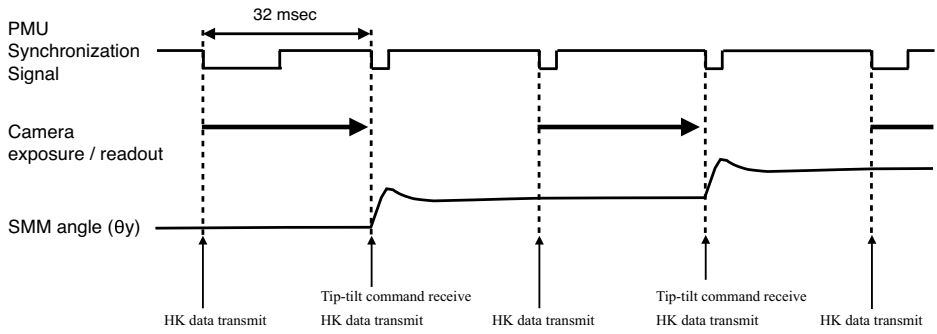}}
\caption{Schematic of the relation between the commands (transmission of HK data and reception of tip-tilt command) and the exposure time of the camera. 
Note that this is the case of the fastest scan mode, where the camera exposure and readout are completed in 32 ms. }
\label{fig:sync}
\end{figure}

Figure \ref{fig:sync} gives an illustrative explanation of the time sequence between the tip-tilt and camera exposure during the scan, in response to the synchronization signal. 
The SCIP-E transmits a command to the SMM every 32 ms, and its timing is specified by the falling edge of the synchronization signal generated by the PMU. 
The SMM tilts the mirror when receiving a tip-tilt command, while returning its HK status to the SCIP-E every 32 ms. 
The SMM adopts a stepping scan system, allocating two types of durations at each scan step, i.e., non-exposure and exposure durations. 
In the non-exposure duration, the tip-tilt angle moves toward the next scan step position. 
In the exposure duration, the camera sensors are activated to be exposed to the incoming light. 
The duration of the non-exposure time equals the interval between a series of synchronization pulses, namely 32 ms.
The duration of the exposure time is adjustable depending on the accumulation of the number of photons. 
In the case of the fastest mode, the interval between scan steps is 64 ms, divided equally between the non-exposure and the exposure times for 32 ms each. 

\subsection{Specifications}
The specifications of the SMM are summarized in Table \ref{SMM-specification}. 
A step resolution of $0.0936^{\prime \prime}$ in the sky angle corresponds to the slit width of the SCIP, which is roughly half of the spatial resolution ($0.21^{\prime \prime}$). 
The scan range of $\pm 33^{\prime \prime}$ covers the FOV of $\pm 29^{\prime \prime}$ well enough to secure a margin in case of misalignment during the assembly process or balloon launch. 
The stability of angle positioning is $3\sigma=0.033^{\prime \prime}$ in the sky angle, where $\sigma$ denotes the standard deviation. 
This requirement is derived from the wavefront error budget to retain the diffraction limited performance \citep{Tsuzuki2020}. 
Mechanical angles are obtained by multiplying the above-mentioned sky angle by the optical magnification (30.52), i.e., the step resolution is converted to 2.857$^{\prime \prime}$, the scan range is $\pm$1005$^{\prime \prime}$, and the stability of the angle positioning is 1$^{\prime \prime}$ ($3\sigma$). 
Note that the optical magnification and the mechanical step resolution is determined considering the fact that the tilt axes created by the flexural pivots have an offset of 25 mm from the mirror surface. 
A linearity error of 0.15$\%$ corresponds to a deviation of one-step divided by the full drive-range of the scan, i.e., $0.0936^{\prime \prime} / (33 ^{\prime \prime} \times 2)=0.00142$. 
The step speed of 32 ms is defined by the time interval between the initiation of the tip-tilt motion and the activation timing of the camera exposure. 
The above mentioned specification enables the extremely fast scan, which completes scanning of the entire range of $\pm 33^{\prime \prime}$ in increments of $0.0936^{\prime \prime}$ (i.e., a total of 705 slit positions) in 39.8 sec, where the angles are expressed in sky angle. 
Whereas a filtergraph provides a similar FOV and time cadence, the slit-based spectrometer offers much wider spectral coverage, including multiple spectral lines. \\
Furthermore, the SMM is to be operated in a thermal and vacuum environment during the flight and ascent phase of the balloon. 
The SMM should be operated at an atmospheric pressure of less than 133 Pa (1 Torr). 
The SMM should maintain its healthiness in the survival temperature range when the SMM is turned off and in the operational temperature range when the SMM is turned on. 
These temperature ranges are separately defined for the SMM-TM/-IF and SMM-DRV (Table \ref{SMM-temp}), because the SMM-TM/-IF should be installed in the ISLiD while the SMM-DRV should be installed on the gondola E-box rack. 

\begin{table}
\caption{Key specifications of the SMM. Angle is described in mechanical angle, whereas angles in parenthesis are in sky angle.}
\label{SMM-specification}
\begin{tabular}{ll}     
  \hline                   
Item & Specification\\
  \hline
Step resolution & 2.857$^{\prime \prime}$ (0.0936$^{\prime \prime}$) \\
Scan range & $\pm$1005$^{\prime \prime}$ ($\pm$32.9$^{\prime \prime}$) \\
Stability (3$\sigma$)& 1.0$^{\prime \prime}$ (0.033$^{\prime \prime}$) \\
Linearity & $<$0.15$\%$ \\
Step speed & $<$32 ms \\
\hline
\end{tabular}
\end{table}

\begin{table}
\caption{Survival and operational temperature ranges of the SMM components ($^\circ C$).}
\label{SMM-temp}
\begin{tabular}{lcccc}     
  \hline                   
 & \multicolumn{2}{c}{Survival} & \multicolumn{2}{c}{Operational} \\
 & lower & higher & lower & higher\\
  \hline
SMM-TM and SMM-IF & -10 & 45 & 0 & 45 \\
SMM-DRV & -20 & 50 & 0 & 45\\
 \hline
\end{tabular}
\end{table}

\section{Optical measurements of SMM performance} 
      \label{S-general}      

\subsection{Setup} 
  \label{S-text}
\begin{figure}    
\centerline{\includegraphics[width=1.0\linewidth,clip=]{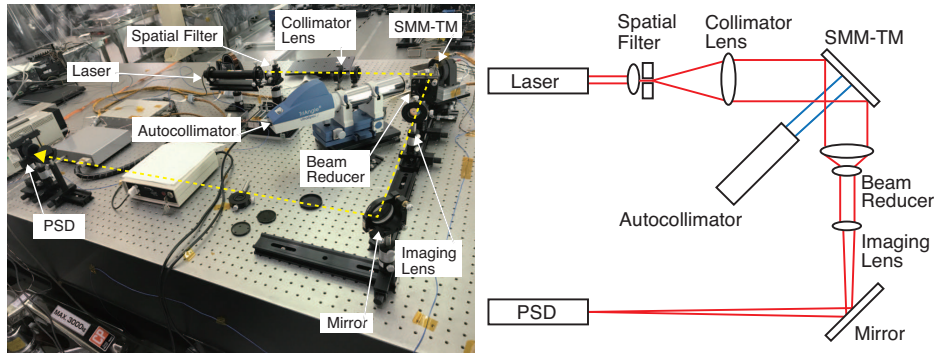}}
\caption{Left: Optical measurement configuration in the performance tests. 
The yellow dotted arrow marks the ray's path to the PSD. Right: Schematic of the optical layout for the autocollimator and the PSD, whose optical rays are shown in blue and red, respectively. }
\label{fig:optical_test}
\end{figure}

An optical measurement test examines the fundamental performance of the SMM, i.e., linearity across the scan range, stability, and step speed. 
We used mainly two types of optical measurement equipment. 
One is an optoelectric autocollimator (TAUS300-57, TRIOPTICS). 
This instrument is suited to measuring the linearity across the full scan range since we can measure the absolute mirror angle. 
The other optical measurement equipment is a position sensitive detector, PSD (S1880, Hamamatsu Photonics). 
The PSD outputs the lateral position of the laser spot illuminated on its detector plane. 
The PSD returns the output at high speed, which is suited for tracing the rapid movement of the scan step and the jittering fluctuation with good temporal resolution. 
The analog signal output by the PSD is read out by the data logger (GL980). \\
The optical test setup is configured as shown in Fig. \ref{fig:optical_test}. 
The autocollimator is set to normally face the mirror, and its optical path is drawn in blue. 
The PSD is set to detect the position of the laser spot displaced by a tilt of the SMM-TM mirror through an afocal system, and its optical path is drawn in red. 
The ray emitted from the laser diverges after passing through a spatial filter that consists of a pin-hole and a lens attached immediately behind the pin-hole. 
The ray, in turn, is collimated by the collimator lens. 
The collimated beam illuminates almost all the effective area of the flat mirror mounted on the SMM-TM, and is reflected at an incident angle of 45 degrees. 
Then the beam diameter is demagnified by the beam reducer. 
Finally, the ray is focused by the imaging lens onto the detector plane of the PSD. 
The baffle is attached immediately in front of the detector plane of the PSD, thereby shutting out unwanted light that enters into the detector. 
All the optical components are arranged on a floating optical bench that efficiently suppresses low-frequency disturbances. \\
In the detector plane of the PSD, the laser position is laterally displaced in accordance with the tip-tilt angle of the SMM-TM. 
The tilt angle is simply given by the equation, $\theta_{y} = d/l$, where $\theta_{y}$ is the tilt angle of the SMM-TM in the scan direction, $d$ is the lateral displacement on the PSD plane, and $l$ is the effective focal length.
The effective focal length is defined by the focal length of the lens (1500 mm) times the magnification of the beam reducer, i.e., three, and thus, $l$ is 4500 mm. 
It should be noted that throughout this section, the mirror angle is expressed in mechanical angle, and not the sky angle. 

\subsection{Results} 
  \label{S-text}
\subsubsection{Linearity} 
  \label{S-text}
\begin{figure} 
 \centering
\includegraphics[width=1.0\linewidth, clip]{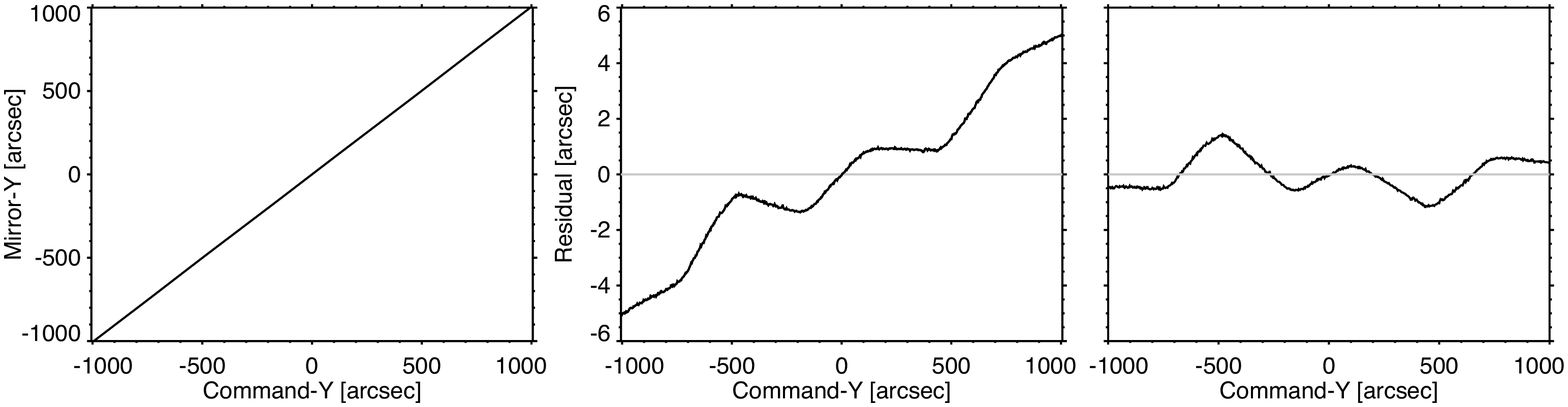} 
\caption{Linearity performance across the full range ($\pm 1005^{\prime \prime}$). 
Left: Mirror angle measured with the autocollimator ($\theta_{measured}$) as a function of the commanded angle ($\theta_{command}$). 
Middle: Residual, calculated by a subtraction of $\theta_{command}$ from $\theta_{measured}$. 
Right: Residual, calculated by a subtraction of the best-fit angle ($\theta_{fit}$) from $\theta_{measured}$. }
 \label{fig:lin}
  \end{figure}

\begin{figure} 
 \centering
\includegraphics[width=1.0\linewidth, clip]{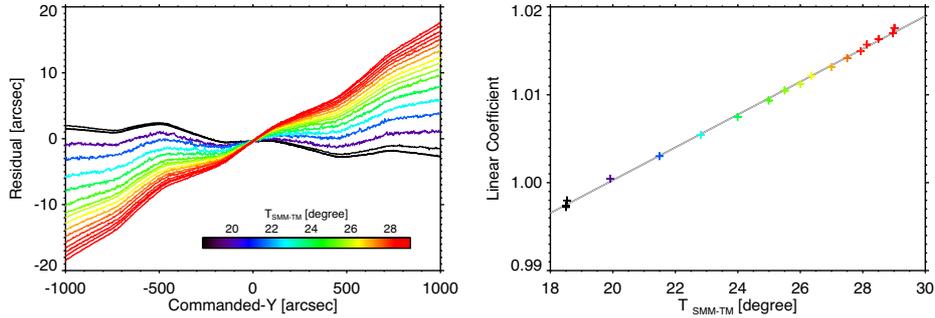} 
\caption{$T_{SMM-TM}$ dependence of a linearity coefficient. 
Left panel is residual between $\theta_{measured}$ and $\theta_{command}$, calculated in the same way as the middle panel of Fig. \ref{fig:lin}. 
The inserted color bar identifies $T_{SMM-TM}$ with that of plots. 
Right panel is the linear coefficient as a function of $T_{SMM-TM}$. 
The color of each cross is consistent with the notation of the left panel. 
The grey solid line is the regression line with coefficients 0.001867 and 0.96292 for the slope and intercept, respectively. }
 \label{fig:lin_temp}
  \end{figure}

The autocollimator is used for measuring the absolute angle of the mirror position. 
The SMM is instructed to scan the entire range of $\pm$1005$^{\prime \prime}$ with a constant spacing of $2.857^{\prime \prime}$, resulting in a total of 705 steps. 
The left panel of Fig. \ref{fig:lin} shows the measured tilt angle, $\theta_{measured}$ with respect to the command angle, $\theta_{command}$. 
It should be noted that $\theta_{measured}$ is offset so that it is $0^{\prime \prime}$ at $\theta_{command}=0^{\prime \prime}$. 
The middle panel of Fig. \ref{fig:lin} displays the residual angle, calculated by subtraction of $\theta_{command}$ from $\theta_{measured}$, presenting a deviation from the ideal scan position depicted by the horizontal grey line. 
The deviation is categorized into two features, i.e., the slope and the non-linearity response. 
For the evaluation of the slope, hereinafter referred to as \textit{the linear coefficient}, $\theta_{measured}$ over the full scan range is fitted with the regression line, 

\begin{equation}
\theta_{fit}= a \theta_{command} + b
\label{lin_fit}
\end{equation}

where $a$ is the linear coefficient and $b$ is an offset parameter. 
In the case of $\theta_{measured}$ in Fig. \ref{fig:lin}, the optimally-fitted parameters are $a=1.0041$ and $b=0.91$. 
The linear coefficient is a measure of the sparsity or density of the scan stepping, e.g., $a=1.0041$ indicates a scanned-range increased by 0.4\% of its original stepping size. \\
The non-linearity is seen as a ripple-like trend with a scale of $500^{\prime \prime}$ across the scan range. 
This means that the step interval is not exactly constant but varies periodically over a scale of $500^{\prime \prime}$. 
The ripple-like trend is primarily caused by the characteristics of the gap-based capacitance sensors because the same behaviour was visible in the measurement of the sensor alone. 
For the evaluation of the non-linearity, the following index is introduced, 

\begin{equation}
Linearity\ error = \frac{ Max( \mid \theta_{measured} - \theta_{fit} \mid) } {1005.66 \times 2} \times 100
\label{eq:lin_err}
\end{equation}

where the operator $Max$ picks up the maximum value among all $\mid \theta_{measured} - \theta_{fit} \mid$ in the 705 steps.  
 $\theta_{measured} - \theta_{fit}$ is plotted in the right panel of Fig. \ref{fig:lin}. 
The maximum of the ripple trend is 1.407$^{\prime \prime}$ at $\theta_{command}=-480^{\prime\prime}$, which corresponds to a linearity error of 0.070\% using equation \ref{eq:lin_err}. 
This value satisfies the requirement of $<$0.2\%, verifying that the SMM covers the full range with an almost constant spacing. \\ 
We also examined the temperature dependence of the linear coefficient. 
The parameter $T_{SMM-TM}$ is the temperature measured with a temperature transducer attached to the SMM-TM, which is one of the HK outputs. 
$T_{SMM-TM}$ is changed by adjusting the temperature of the experiment room using an air conditioner. 
The left panel of Fig. \ref{fig:lin_temp} shows the residual of the middle panel of Fig. \ref{fig:lin} at different $T_{SMM-TM}$. 
$T_{SMM-TM}$ dependence of the offset $b$ in equation \ref{lin_fit}, is adjusted so that all the plots coincide at $\theta_{command}=0^{\prime \prime}$. 
It is clear that large $T_{SMM-TM}$ increases the interval of the scan step across the full FOV, whereas small $T_{SMM-TM}$ narrows the interval of the scan step which overlap each other slightly. 
It is of note that the linear coefficient is well correlated with $T_{SMM-TM}$, and can be expressed by a regression line with a slope of 0.001867 and an intercept of 0.96292. 
The standard deviation of their residual is only 0.0002454, indicating that they are tightly correlated.
The $T_{SMM-TM}$ dependence of the step interval can be ascribed to the thermal expansion of the SMM-TM material and the temperature dependence of the signal output by the capacitance gap sensors. 
For application to observation data, post processing is done to implement a correction procedure for this temperature dependence based on $T_{SMM-TM}$. 
In contrast, the linearity error falls within a range of $0.07\%-0.08\%$, irrespective of $T_{SMM-TM}$ over 19-29$^\circ C$, fulfilling the above requirements. 

\subsubsection{Jitter} 
\begin{figure} 
 \centering
\includegraphics[width=1.0\linewidth, clip]{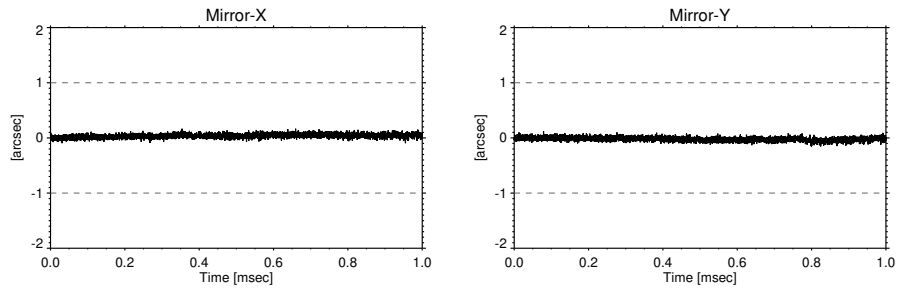} 
\caption{Jitter components of $\theta_{x}$ and $\theta_{y}$. 
Horizontal grey dashed line marks the requirement of $3 \sigma=\pm 1.0^{\prime \prime}$. 
Note that mirror angle is expressed in mechanical angle. }
 \label{fig:jit}
  \end{figure}

The PSD gives jittering fluctuation of the mirror with a sampling rate of 0.1 ms. 
Fig. \ref{fig:jit} shows the time variation of $\theta_{x}$ and $\theta_{y}$. 
The amplitude of the fluctuation in $\theta_{x}$ and $\theta_{y}$ are 0.09$^{\prime \prime}$ and 0.10$^{\prime \prime}$ ($3\sigma$), respectively. 
It should be noted that these angles are expressed in mechanical angle, e.g., $0.1^{\prime \prime}$ in mechanical angle is converted to $0.0033^{\prime \prime}$ in sky angle for reference. 
Irrespective of the oriented angle in the FOV, jitter remains almost constant between 0.08$^{\prime \prime}$ and 0.10$^{\prime \prime}$. 
Table \ref{tab:jit} summarizes the results at the angle position close to the scan edges ($\theta_{y}=\pm 793^{\prime \prime}$) and the offsets in the perpendicular direction ($\theta_{x}=\pm 201^{\prime \prime}$). 
$\theta_{x}$ is only decenterd by $\pm 201^{\prime \prime}$ since this angle is defined as the offset tolerance used for the coalignment with other observation instruments aboard \textsc{Sunrise III}. 
As a result, the jitter components of $\theta_{x}$ and $\theta_{y}$ over the full FOV are nearly one-order lower than the requirement of $1.0^{\prime \prime}$ ($3\sigma$). 

\begin{table}
\caption{FOV dependence of jitter. Two values at each blank denote $\theta_{x}$ and $\theta_{y}$ in left and right, respectively. }
\label{tab:jit}
\begin{tabular}{cccc} 
\hline                   
 & $\theta_{x}$=-201$^{\prime \prime}$ & $\theta_{x}$=0$^{\prime \prime}$ & $\theta_{x}$= 201$^{\prime \prime}$\\ \hline
$\theta_{y}$=793$^{\prime \prime}$ & 0.10$^{\prime \prime}$, 0.08$^{\prime \prime}$ & 0.09$^{\prime \prime}$, 0.09$^{\prime \prime}$ & 0.09$^{\prime \prime}$, 0.09$^{\prime \prime}$ \\
$\theta_{y}$=0$^{\prime \prime}$ & 0.09$^{\prime \prime}$, 0.09$^{\prime \prime}$ & 0.10$^{\prime \prime}$, 0.09$^{\prime \prime}$ & 0.09$^{\prime \prime}$, 0.09$^{\prime \prime}$ \\
$\theta_{y}$=-793$^{\prime \prime}$ & 0.09$^{\prime \prime}$, 0.09$^{\prime \prime}$ & 0.09$^{\prime \prime}$, 0.09$^{\prime \prime}$ & 0.09$^{\prime \prime}$, 0.09$^{\prime \prime}$\\ \hline
\end{tabular}
\end{table}

\subsubsection{Step speed} 
\begin{figure} 
 \centering
\includegraphics[width=1.0\linewidth, clip]{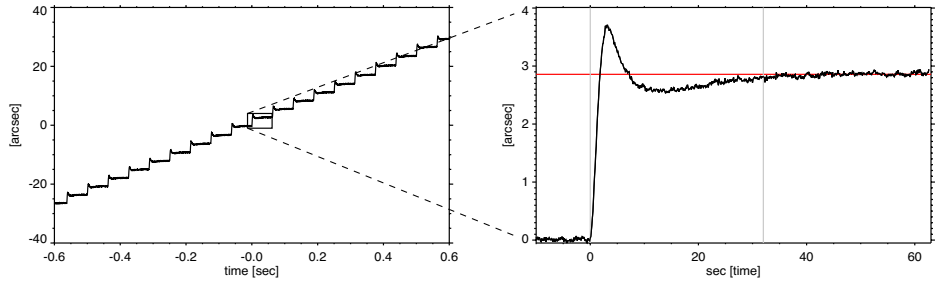} 
\caption{Left: Mirror angle of $\theta_{y}$ during the fastest scan mode (64 ms step interval), measured by the PSD at a sampling rate of 0.1 ms. Right: Enlarged view of the mirror angle for one step from 0$^{\prime \prime}$ to 2.857$^{\prime \prime}$. 
Time axis is adjusted so that 0 ms corresponds to the start time of the stepping from 0$^{\prime \prime}$ to 2.857$^{\prime \prime}$. 
Vertical grey line marks 32 ms after the step start. 
Horizontal red line indicates the adjacent step position ($\theta_{y}=2.857^{\prime \prime}$). }
 \label{fig:step_center}
  \end{figure}

\begin{figure} 
 \centering
\includegraphics[width=1.0\linewidth, clip]{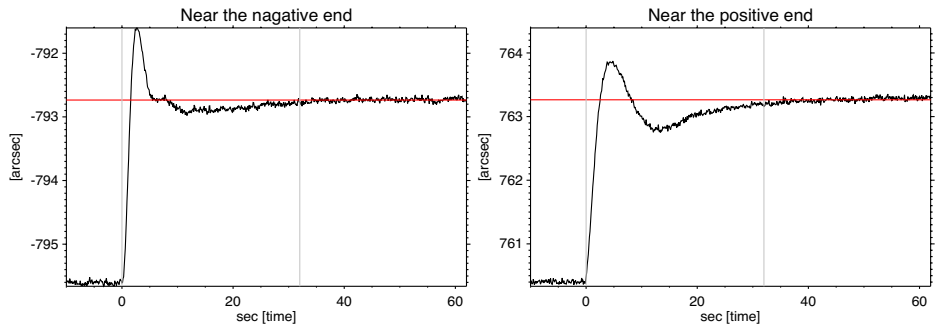} 
\caption{Same plots as the right panel of Fig.\ref{fig:step_center} but for one step from around $-796^{\prime \prime}$ and $+760^{\prime \prime}$ in the left and right panels, respectively. }
 \label{fig:step_minus_plus}
  \end{figure}

In a scan stepping, the mirror angle should move and stabilize at the adjacent step position within 32 ms. 
Fig. \ref{fig:step_center} shows a time sequence of the scan stepping near the center of the FOV, measured with the PSD at a sampling rate of 0.1 ms. 
The right panel of Fig. \ref{fig:step_center} presents a closer look at one step behavior. 
The mirror angle quickly rises with overshoot, falls down with undershoot, and finally approaches the commanded angle steadily. 
The arrival time is defined as when the mirror angle reaches 2.757$^{\prime \prime}$, which is determined by the commanded position 2.857$^{\prime \prime}$ minus the jitter component 0.1$^{\prime \prime}$.
The arrival time is 26.1 ms on average, fulfilling the requirement of being smaller than 32 ms. \\
The arrival time depends on the FOV. 
The left panel of Fig. \ref{fig:step_minus_plus} shows the temporal behavior near the negative end, corresponding to the start position of a scan. 
The mirror angle overshoots sharply and, in turn, undershoots slightly, shortening the arrival time to 21 ms. 
The right panel of Fig. \ref{fig:step_minus_plus} shows the same,  but near the positive end, corresponding to the end position of a scan. 
The mirror angle overshoots mildly and then undershoots largely, prolonging the arrival time to 27 ms. 
Thus, the arrival time at both ends of the scan fulfils the requirement of being within 32 ms. 
The FOV dependence is due to slightly unequal distances between the coils and the metal plates in the electromagnetic actuators.
This induces a small imbalance in the strength of the magnetic pulling force. 
In contrast, this imbalanced force should not affect the arrival position because the closed-loop system controls the mirror angle toward the arrival position. 

\subsubsection{Flyback motion} 
\begin{figure} 
 \centering
\includegraphics[width=1.0\linewidth, clip]{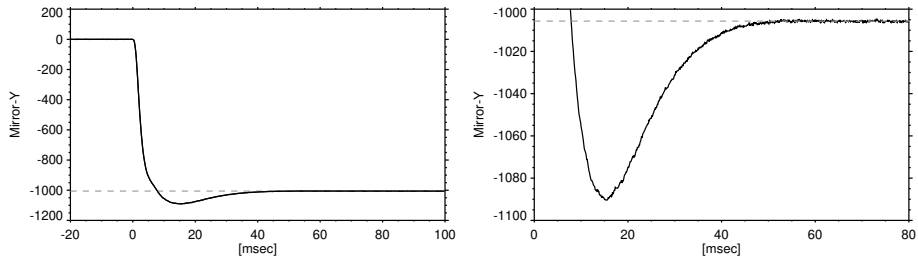}
\caption{Mirror angle in the flyback movement from 0$^{\prime \prime}$ to -1005.66$^{\prime \prime}$. 
The horizontal grey dashed line marks the position of -1005.66$^{\prime \prime}$ (the starting position of the scan). 
Time axis is adjusted so that 0 ms coincides with the starting  time of the flyback movement. 
The overall movement is described in the left panel, and its magnified view around -1005.66$^{\prime \prime}$ is shown in the right panel.}
 \label{fig:flyback}
  \end{figure}

In the repetition of the scan, the flyback motion is a necessary process as it returns from the end position to the starting position. 
The flyback time should be minimized, particularly for repetitive scans with high temporal cadence, since no science observations can be made during that time. 
The optical measurement test was performed to investigate the duration of the flyback motion.
This test is similar to that of the step-speed measurement (section 3.2.3) but the difference 
is that a large angle motion, which is oppositely directed to the nominal scan, is tested.
The configuration setup is similar to the step-speed measurement (Fig. \ref{fig:optical_test}). 
The exception is that the imaging lens was replaced with another lens whose focal length is 30 cm, and the PSD was placed at the focused position accordingly. 
This modification of the optical configuration is to shorten the effective focal length, thereby making displacement of the laser spot over the entire angle range fall within the limited detector area of the PSD. 
Fig. \ref{fig:flyback} depicts a flyback motion ranging from 0$^{\prime \prime}$ to -1005.66$^{\prime \prime}$. 
The large overshoot associated with the flyback reaches -1090$^{\prime \prime}$. 
The mirror angle in this motion takes only 50 ms to reach the starting position (-1005.66$^{\prime \prime}$). 
Table \ref{tab:flyback} shows the flyback time and the overshoot for various flyback spans. 
The flyback time remains almost constant within a range of 50 to 57 ms, although the flyback span differs by more than one order of magnitude. 
In contrast, the overshoot drastically increases with the widening of the flyback span. 
These characteristics are consistent with the inherent aspect of the PID control implemented on the closed-loop system, i.e., an overshoot increases while retaining almost the same arrival time.  \\
One concern is a situation where an overshoot largely beyond the edge of the FOV hits the mechanical limit of the SMM and damages it. 
To prohibit an overshoot motion from hitting, a mechanical stopper is set at a distance beyond the edges (roughly $\pm$1300$^{\prime \prime}$).  
Furthermore, to prevent even hitting this stopper, we implemented a function that inserts an intermediate step into the flyback procedure, in which the mirror angle moves to an angle between the end and start positions once, and then moves to the start position. 
Thus, the intermediate step function divides a flyback into two steps and reduces the span by half at one flyback, thereby effectively suppressing the overshoot at the cost of doubling the flyback duration. 
In a comparison between overshoot with and without the intermediate step, for example, the former reaches only -1090$^{\prime \prime}$ (Fig. \ref{fig:flyback}) while the latter reaches -1299$^{\prime \prime}$ (bottom row of Table \ref{tab:flyback}). \\
Regardless of the flyback span, 128 ms including a sufficient margin is allotted to one flyback time. 
In the case of inserting the intermediate step, a total of 256 ms is allotted. 
Interruption of such a short time accounts for only 0.3-0.6\% even at the case of the highest cadence of 40 sec for the full scan ($\pm 1005^{\prime \prime}$). 

  \begin{table}
\caption{Flyback times and the corresponding overshoot for various flyback spans.}
\label{tab:flyback}
\begin{tabular}{rclcc}     
  \hline                   
 & flyback time [ms] & overshoot [arcsec] \\
  \hline
$+30^{\prime \prime}$   to $-30^{\prime \prime}$ & 52.3 & $-60.8^{\prime \prime}$ \\
$+60^{\prime \prime}$   to $-60^{\prime \prime}$ & 53.3 & $-102.0^{\prime \prime}$ \\
$+200^{\prime \prime}$  to $-200^{\prime \prime}$ & 52.0 & $-204.0^{\prime \prime}$ \\
$+300^{\prime \prime}$  to $-300^{\prime \prime}$ & 50.3 & $-323.7^{\prime \prime}$ \\
$+400^{\prime \prime}$  to $-400^{\prime \prime}$ & 56.8 & $-448.1^{\prime \prime}$ \\
$+500^{\prime \prime}$  to $-500^{\prime \prime}$ & 54.5 & $-570.2^{\prime \prime}$ \\
$+800^{\prime \prime}$  to $-800^{\prime \prime}$ & 56.1 & $-962.3^{\prime \prime}$ \\
$+1000^{\prime \prime}$ to $-1000^{\prime \prime}$ & 54.8 & $-1299.0^{\prime \prime}$ \\
 \hline
\end{tabular}
\end{table}

\subsubsection{Long run} 
The SMM is to be operated during almost the whole period of the ballon flight. 
A long-run test was accordingly executed for a total of 120 hours, corresponding to the expected science observation period of 5 days. 
The duration of the long-run test was divided into 4 types under different operational conditions: 
40 hours for 5040 cycles of enabling and disabling the closed-loop control; 42 hours for spectropolarimetric scan mode with an exposure time of 512 ms at each step; 14 hours for spectroscopic scan mode with the shortest exposure of 10 ms, and 24 hours for being simply left with the closed-loop control enabled. 
The total number of repeated scans in these spectroscopic and spectropolarimetric modes were 950 and 335 scans, which are equivalent to 670,000 and 236,510 steps, respectively. 
No malfunction and no degradation in the linearity performance were reported during the test. 
Thus the long-run test ensured that the SMM retains its functions throughout the entire period of the science operation.

\subsection{Environmental Tests} 
\label{S-labels}

\subsubsection{Thermal-Cycle Test} 
\begin{figure}    
\centerline{\includegraphics[width=0.9\linewidth,clip=]{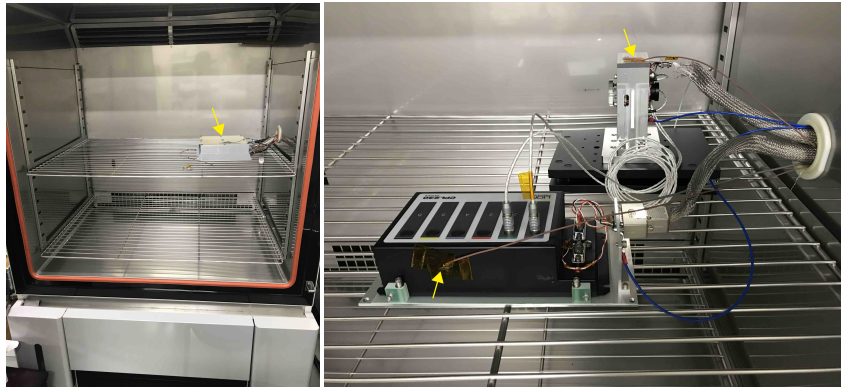}}
\caption{Photograph of the SMM-DRV, and SMM-TM together with SMM-IF, placed inside the thermal chamber with its door opened. 
Yellow arrow indicates the location of the attached thermocouples to monitor the temperatures of each SMM component. }
\label{fig:thermal_vac_photo}
\end{figure}

\begin{figure}    
\centerline{\includegraphics[width=0.9\linewidth,clip=]{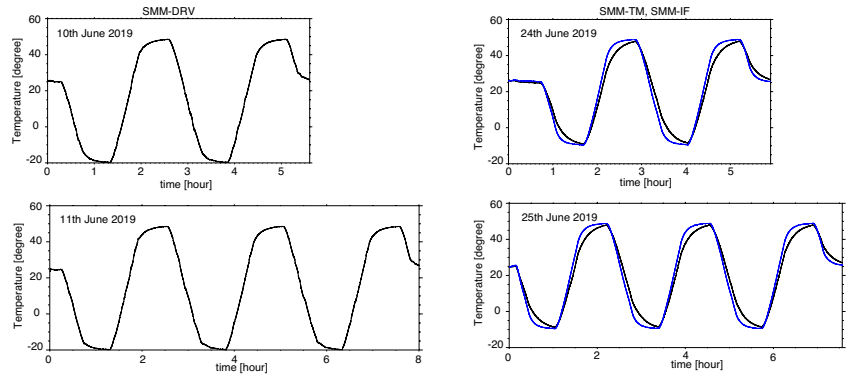}}
\caption{Temperature of the SMM-DRV and SMM-TM (black) together with SMM-IF (blue), during the thermal cycles in the left and the right columns, respectively. 
The length of horizontal axes between the 1st and 2nd date are adjusted so that the scale of the time is same. }
\label{fig:thermal}
\end{figure}

The SMM is to be operated in a balloon-flight environment. 
Heat load is associated with the ascent phase and the flight, and it may induce equipment failure. 
The thermal-cycle test is accordingly executed to verify that the SMM maintains its healthiness after being exposed to such a temperature variation. 
The SMM-TM and SMM-IF were placed together inside a thermal chamber (Fig. \ref{fig:thermal_vac_photo}) and temperature-load variation from -10$^\circ C$ to +50$^\circ C$ was applied. 
A temperature variation of -20$^\circ C$ to +50$^\circ C$ was separately applied to the SMM-DRV, since the survival temperature of the SMM-DRV differs from that of the SMM-TM and SMM-IF (Table. \ref{SMM-temp}). 
A total of 5 cycles of the temperature variation were loaded with the SMM components powered off. 
Thermocouples are attached to all the SMM components in order to monitor their temperature since no HK is transmitted when the SMM is turned off. 
In the recorded temperature (Fig. \ref{fig:thermal}), the minimum and maximum temperatures reached -9.5$^{\circ}C$ and 48.9$^{\circ}C$, respectively, for the SMM-TM and SMM-IF, whereas those of the SMM-DRV reached -19.9$^{\circ}C$ and 48.6$^{\circ}C$, respectively. 
After loading the temperature cycles, we powered on the SMM and successfully executed its scan function at the highest and lowest operational temperatures. 
This thermal cycle test demonstrates the SMM meets the requirements for the range of the survival and operational temperature ranges.

\subsubsection{Thermal-Vacuum Test} 
\begin{figure}    
\centerline{\includegraphics[width=1.0\linewidth,clip=]{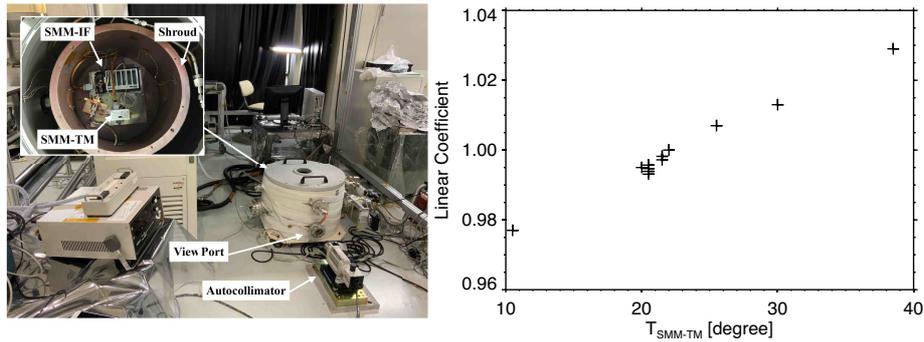}}
\caption{Left: Overview of the thermal vacuum test environment. 
Insert shows the inside of the vacuum chamber with the top panel of the shroud opened. 
Right: $T_{SMM-TM}$ dependence of the linear coefficient under the vacuum condition. }
\label{fig:thermal_vac}
\end{figure}

The thermal-vacuum test is to validate that the SMM is properly operated in the vacuum environment, together with low and high temperature conditions. 
The SMM-TM and SMM-IF were placed inside a vacuum chamber (left panel of Fig. \ref{fig:thermal_vac}), in which the ambient pressure was maintained at around $10^{-2}$ Pa. 
Inside the chamber, the SMM components were connected to the bottom of the copper shroud so that their temperatures could be easily changed by circulating a temperature-controlled refrigerant into the shroud. 
In the test, we turned on the SMM at temperatures of 10, 20, 25, 30, 45$^{\circ}C$, and then confirmed that the SMM successfully executed its scan function. 
To evaluate the linear coefficient, another optoelectric autocollimator (PA102, Nikon) was placed immediately outside the vacuum chamber in front of the SMM-TM through the view port of the chamber. 
The right panel of Fig. \ref{fig:thermal_vac} shows the linear coefficients measured at several $T_{SMM-TM}$. 
It is clear that the linear coefficient in the vacuum condition increases linearly with increasing $T_{SMM-TM}$ through a range of 10 - 40$^{\circ}C$. 
The linear coefficients at 20 - 30$^{\circ}C$ were confirmed to be consistent with the results measured at atmospheric pressure (Fig. \ref{fig:lin_temp}). 
We evaluated the jitter by analyzing the mirror angle data recorded in HK rather than that measured by the optoelectric measurement device, since its accuracy is poor when the optical path passes through its view port. 
The resultant jitter is in the range 0.1$^{\prime \prime}$-0.4$^{\prime \prime}$ (3$\sigma$), which is significantly better than the $1.0^{\prime \prime}$ requirement. 
Some of the jitter values are larger than those measured at atmospheric pressure (0.08$^{\prime \prime}$-0.10$^{\prime \prime}$ in Table. \ref{tab:jit}). 
This can be attributed to mechanical disturbances transmitted from the vacuum chamber and the ground, because we did not use a vibration-isolated table in the test. 

\subsection{Synchronization Tests} 
\label{S-labels}
\subsubsection{Response time} 
\begin{figure}    
\centerline{\includegraphics[width=1.0\linewidth,clip=]{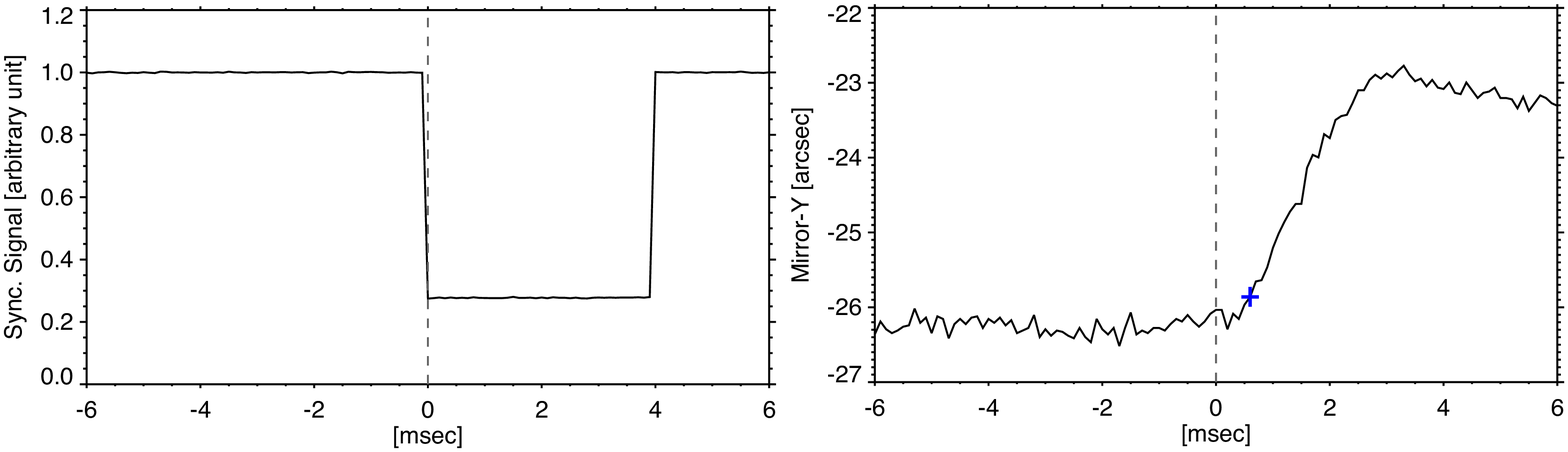}}
\caption{Synchronization signal and the tip-tilt angle, whose analog signals are recorded with the same data logger, are shown in the left and right panels, respectively.
The data logger produces the analog outputs at a sampling rate of 0.1 ms. 
Horizontal axis gives the elapsed time from the falling edge in the synchronization signal (emphasized by the vertical dashed line). 
The blue cross marks a response time determined by our definition. }
\label{fig:synctest1}
\end{figure}

The SMM executes the scan function together with the other electronics, since a tip-tilt command is transmitted from the SCIP-E and its timing is specified by the synchronization signal generated by the PMU (section 2.1). 
The mirror movement should be initiated with a minimal time lag after receiving a tip-tilt command. 
This time lag is hereinafter referred to as \textit{the response time}. 
The purpose of the synchronization test is to inspect whether the response time meets the requirement of being less than 5 ms, derived as the subtraction of the step speed (27 ms) from the allotted duration for one scan step (32 ms). \\
The test setup is configured as shown in Fig. \ref{fig:optical_test}. 
The exception is that the analog line of the synchronization signal from the PMU is set to be read out by a data logger together with the output of the PSD so that the logger records both signals with an exactly identical time-stamp. 
Fig. \ref{fig:synctest1} shows the time evolution of the mirror movement in response to the analog signal of the synchronization pulse. 
The timing of the tip-tilt command is specified by the falling edge of the synchronization pulse. 
The response time is defined as the time taken by $\theta_{measured}$ to deviate by more than 3$\sigma$, where $\sigma$ denotes a standard deviation of $\theta_{measured}$ over the period from -6 to 0 ms. 
In the case of Fig. \ref{fig:synctest1}, the mirror movement lags by 0.6 ms with respect to the falling edge of the synchronization signal. 
Twenty two other examples were analyzed in the same manner, and their results are in the range $0.52\pm$0.13 ms. 
This value verifies that the SMM satisfies the requirement for the response time. \\

\subsubsection{Camera synchronization} 
\begin{figure}    
\centerline{\includegraphics[width=1.0\linewidth,clip=]{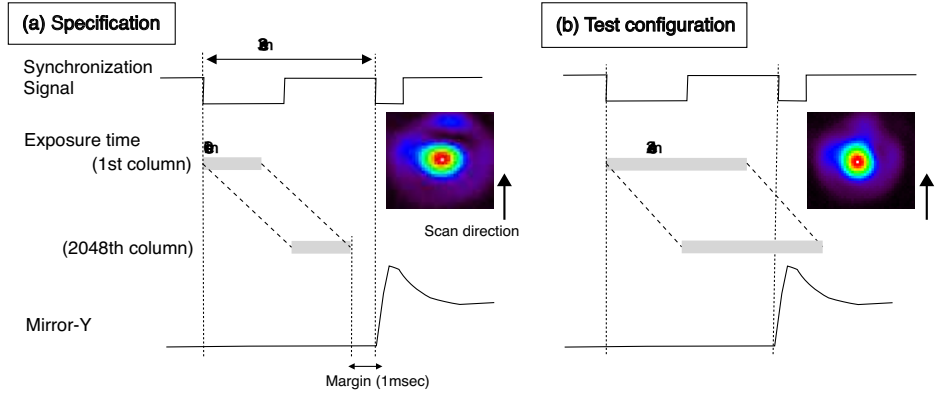}}
\caption{Relation between the exposure time at each column of the CMOS sensors and mirror movement, in response to the synchronization signal. 
Exposure time is shown by the grey rectangular box (explicitly shown only at the top and bottom columns).
The sensors are exposed sequentially from the top to bottom columns, as depicted by the inclined dashed lines. 
The inserted images are the actually captured laser spots. 
Case (a): Fastest scan mode with an exposure time of 10 ms. 
Case (b): Test mode with an exposure time of 24 ms. 
The laser spot with the exposure time of 24 ms is elongated in the scan direction, compared with the original one with exposure time of 10 ms. }
\label{fig:synctest2_concept}
\end{figure}

\begin{figure}    
\centerline{\includegraphics[width=1.0\linewidth,clip=]{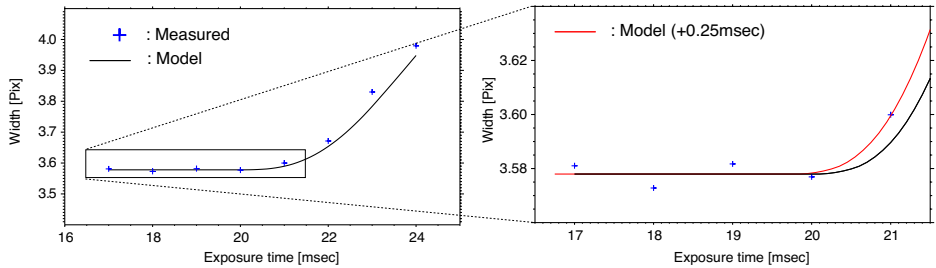}}
\caption{Left: Measured width of the spot centered at the 995th column at different exposure times (blue cross), together with the model calculation (black line). Right: Close up view of the left panel, especially enlarging the rising up timing of the width. 
The red line is the plot of the model calculation with a time lag of 0.25 ms with respect to the designed timing of the camera exposure. }
\label{fig:sync2_result}
\end{figure}

The synchronization signal controls not only the timing of the tip-tilt command but also the timing of the camera exposure to be activated 32 ms later (section 2.1). 
This test is accordingly executed to verify the synchronization between the mirror movement and the camera exposure. 
The flight model of the CMOS camera is GSENSE400BSI (GPIXELs Inc.), using a rolling shutter system in which the individual columns of the camera sensor are exposed sequentially, and followed by the readout process that takes place column by column (case $a$ in Fig. \ref{fig:synctest2_concept}). 
It takes 10.2 {\textmu}s to read out one column, resulting in about 21 ms to read out one image consisting of 2048 columns.
In the case of the fastest scan in the SCIP observation, the exposure time takes 10 ms so that the readout process finishes at 31 ms with a margin of 1 ms in order to complete the processes of the exposure and readout within 32 ms safely. \\
For checking the actual behavior of this design, we intentionally used an exposure time longer than 10 ms (case $b$ in Fig. \ref{fig:synctest2_concept}). 
If the exposure time is set to be longer than 11 ms (it uses up the margin of 1 ms), it includes a period of mirror movement near the end column of the sequential process of the exposure, and blurs the obtained image at the columns. 
The optical measurement environment is configured again as shown in Fig. \ref{fig:optical_test}, except for the PSD being replaced with the flight camera. 
In our measurement setup, the laser spot illuminating on the CMOS sensor is centered at the 995th column, where the image blur is expected if the exposure time is set to be longer than 20.2 ms, as calculated by $10 + 0.0102 \times 995$. \\
To confirm whether an exposure time longer than 20.2 ms induces image degradation, we varied the exposure time from 17 to 24 ms in steps of 1 ms. 
The degree of the image blur is evaluated by introducing spot width in the scan direction, which is calculated from the full width half maximum of the Gaussian function optimally fitted to the captured laser spot image. 
The blue crosses in Fig. \ref{fig:sync2_result} depict the measured spot width as a function of the exposure time. 
It can be seen that the image starts to blur at around 20 - 21 ms, and it is consistent with the ideal design of 20.2 ms. \\
For a detailed inspection, the image blur at the 995th column as a function of the exposure time was modelled. 
We introduced a synthesized image in order to compare its spot width with that of the measured image.
The synthesized image was modelled to consist of multiple sub-images segmented in time within the exposure duration. 
This synthesized image $I$ is expressed as  
\begin{equation}
I = \frac{1}{n} \sum_{t=1}^{n} I^{\prime} (t), 
\label{eq:cam}
\end{equation}
where $I^{\prime}(t)$ is the sub-image segmented in time, $n$ is the number of the segments obtained by dividing the exposure duration by 0.1 ms, e.g., if the exposure time is 24 ms, $n$ is 240 and each $I^{\prime}(t)$ is an image at each interval of 0.1 ms. 
Each $I^{\prime}$ at a given time is simply created by shifting the illuminated spot of \textit{a no-motion image} according to the mirror movement (Fig. \ref{fig:step_center}). 
Here \textit{no-motion image} refers to the image actually captured with the shortest exposure time of 10 ms, i.e., the laser spot of panel $a$ in Fig. \ref{fig:synctest2_concept}. 
The spot width of the synthesized image $I$ as a function of exposure time is plotted with the black solid line in Fig. \ref{fig:sync2_result}. 
The red solid line is a plot of the model calculation with a time lag of 0.25 ms with respect to the exposure time, so that the model calculation coincides with the measured spot-width at the exposure time of 21 ms. \\
Thus, the synchronization accuracy that fulfils the requirement for shorter than 1 ms time lag, and does not induce an image degradation was verified. 
The purpose of this synchronization test is similar to the test for response time (section 3.4.1), but the former test inspects the synchronization behavior under the electric end-to-end configuration. 

\section{Integration test} 
      \label{S-features}
\begin{figure} 
\centerline{\includegraphics[width=1.0\linewidth,clip=]{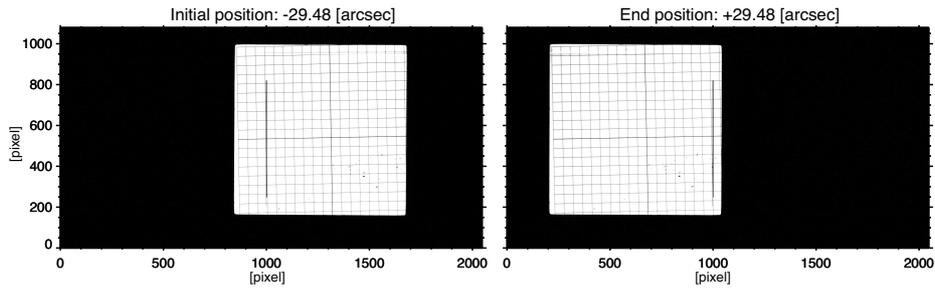}}
\caption{Grid images taken by the slit-jaw camera of the SCIP. 
Left and right panels represent images at the positions of $\theta_{command}=$-29.48$^{\prime \prime}$ and $+29.48^{\prime \prime}$, respectively. 
Only 1024 rows near the central position in the vertical direction are drawn, while the format contains up to 2048 rows. 
The slit is located at the 1000th pixel in the horizontal direction of the camera, spanning between the 250th and 820th pixels in the vertical direction. 
Note that the slit width is artificially broadened for better visualization. }
\label{fig:grid_img}
\end{figure}

\begin{figure} 
\centerline{\includegraphics[width=1.0\linewidth,clip=]{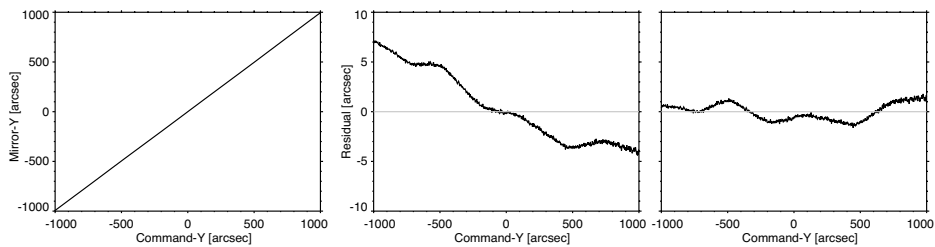}}
\caption{
Linearity performance of the SMM across the full FOV in the state of being fully assembled and installed into the postfocus instrumentation of \textsc{Sunrise III}. 
Left: $\theta_{measured}$ as a function of $\theta_{command}$, where $\theta_{measured}$ is calculated from the corresponding grid position. 
Middle: Residual, calculated by a subtraction of $\theta_{command}$ from $\theta_{measured}$. 
Right: Residual, calculated by a subtraction of $\theta_{fit}$ from $\theta_{measured}$. }
 \label{fig:lin_integ}
\end{figure}

The linearity performance of the SMM was tested in the same configuration as the actual flight, i.e., the SMM was mounted on the ISLiD. 
The SCIP was fully assembled and mounted on the postfocus instrumentation of \textsc{Sunrise III}. 
In this test configuration, the beam illuminates a grid target placed at the secondary focus position in the ISLiD. 
The beam in turn, is fed by the ISLiD optics, including SMM-TM, onto the entrance slit of the SCIP. 
For the evaluation of the linearity performance, we used images taken by the slit-jaw camera of the SCIP rather than the spectrograph cameras. 
The purpose of installing the slit-jaw camera is to provide context by capturing the spatially two-dimensional image reflected at the entrance slit plane outside the slit. 
We executed a full-range scan and recorded a total of 705 grid images taken at each step position (Fig.\ref{fig:grid_img}). 
For all the consecutive images, we traced a grid position moving across the FOV (Fig. \ref{fig:lin_integ}). 
$\theta_{measured}$ in mechanical angle is given by the conversion of 1 pixel to 2.857$^{\prime \prime}$ in the ideal design. 
Although $\theta_{measured}$ no longer meaningfully reflects the mechanical angle, the angle is still expressed in mechanical angle for the consistency throughout the paper. 
The linear coefficient is 0.994 (the middle panel of Fig. \ref{fig:lin_integ}), which is lower than the 1.004 measured in the optical test for the SMM component alone (Fig. \ref{fig:lin_temp} at the corresponding temperature of $T_{SMM-TM}=22.5^{\circ}C$). 
This deviation of 1\% can be attributed to a small error in the curvature radius of the relay optics in the ISLiD. 
An error in the pixel scale in the slit-jaw optics also contributes to the deviation. 
Regarding the non-linearity performance, the ripple-trend is also seen in the right panel of Fig. \ref{fig:lin_integ}. 
The ripple peaks coincide with the measurement result of the optical test for the SMM component alone (Fig. \ref{fig:lin}). 
The maximum non-linearity is 1.81$^{\prime \prime}$, which converts to a linearity error of 0.090\% using the equation (\ref{eq:lin_err}). 
Thus, we confirmed that the SMM maintains good linearity performance when operated in the the same configuration as the actual flight.

\section{Summary} 
      \label{S-features}      
We developed a scan mirror mechanism for a slit-based spectrometer or spectropolarimeter to precisely and quickly map the physical quantities of the solar atmosphere. 
The SMM achieved the following results: 
supreme scan-step uniformity (linearity of 0.08$\%$) across the broad field-of-view ($\pm 1005^{\prime \prime}$), and high stability ($3\sigma =0.1^{\prime \prime}$), where the angles are described in mechanical angle. 
The fast scan speed, which takes only 32 ms from one step position to another, is particularly impressive; for example, the full FOV with a stepping of $2.857^{\prime \prime}$ is scanned within only 40 s. 
A flyback motion for a repeated scan takes only $<$57 ms to return from the end to the start position, regardless of the scan range upto $\pm 1005^{\prime \prime}$. 
These fundamental performances are required for research into magnetohydrodynamics of the photosphere and the chromosphere. \\
In the near future, the capability of the SMM will be demonstrated in actual use as the mechanism is installed for near-infrared spectropolarimetry (SCIP) onboard the balloon-borne solar observatory \textsc{Sunrise III}, which is scheduled to be launched in 2022. 
The SMM has been verified to be compatible with the near space environment since the mechanism is required to operate at severe pressure/temperature conditions during the balloon-flight in the stratosphere. 
The SMM has the potential to be applied in space-borne telescopes in the future with careful selection of its material considering different conditions, such as the radiation resistance.

\begin{acks}
The balloon-borne solar observatory \textsc{Sunrise III} is a mission of the Max Planck Institute for Solar System Research (MPS, Germany), and the Johns Hopkins Applied Physics Laboratory (APL, USA). \textsc{Sunrise III} looks at the Sun from the stratosphere using a 1-meter telescope, three scientific instruments, and an image stabilization system. Significant contributors to the mission are a Spanish consortium, the National Astronomical Observatory of Japan (NAOJ, Japan), and the Leibniz Institute for Solar Physics (KIS, Germany). The Spanish consortium is led by the Instituto de Astrof\'{i}sica de Andaluc\'{i}a (IAA, Spain) and includes the Instituto Nacional de T\'{e}cnica Aeroespacial (INTA), Universitat de Val\`{e}ncia (UV), Universidad Polit\'{e}cnica de Madrid (UPM) and the Instituto de Astrof\'{i}sica de Canarias (IAC). Other partners include NASA's Wallops Flight Facility Balloon Program Office (WFF-BPO) and the Swedish Space Corporation (SSC). \textsc{Sunrise III} is supported by funding from the Max Planck
Foundation, NASA under Grant \#80NSSC18K0934, Spanish FEDER/AEI/MCIU (RTI2018-096886-C5) and a “Center of Excellence Severo Ochoa” award to IAA-CSIC (SEV-2017-0709), and the ISAS/JAXA Small Mission-of-Opportunity program and JSPS KAKENHI JP18H05234, and NAOJ Research Coordination Committee, NINS. We would also thank significant technical support from the Advanced Technology Center (ATC), NAOJ. We would like to thank Editage (www.editage.com) for English language editing.
\end{acks}

\bibliographystyle{spr-mp-sola}
\bibliography{sola_bibliography_example.bib}

\end{article}
\end{document}